# Life history stage effects on alert and flight initiation distances in king penguins (*Aptenodytes patagonicus*)

Tracey L Hammer[1], Pierre Bize[2,3], Benoit Gineste[1], Jean-Patrice Robin[1], René Groscolas[1], Vincent A Viblanc[1]

[1]Université de Strasbourg, CNRS, IPHC UMR 7178, F-67000 Strasbourg, France

[2]Swiss Ornithological Institute, CH-6204 Sempach, Switzerland

[3]Department of Ecology and Evolution, University of Lausanne, Biophore, CH-1015 Lausanne

**ABSTRACT**

When approached by predators, prey must decide whether to flee or remain and fight. The economics of such decisions are underlain by the trade-off between current and residual fitness. The trade-off predicts that (i) breeders should be less prone than non-breeders to flee from approaching predators, as breeders can lose their investment into current reproduction; (ii) among breeders, parents should increasingly defend their offspring with increasing investment into the brood (brood value hypothesis), at least until the offspring can independently take part in anti-predator defenses; and (iii) for a similar investment into reproduction, breeders with lower perspectives to fledge or wean their young should invest less into offspring defense. We tested these predictions in a colonially breeding seabird, the king penguin (*Aptenodytes patagonicus*). Specifically, we considered how antipredator behaviors varied according to life history stage (molting, courting, breeding), offspring age and their dependence on parents for antipredator defenses, and the timing of breeding, with late breeders being very unlikely to fledge offspring in this species. Using non-lethal human approaches to mimic the threat of predation, we approached >500 penguins and measured their alert and flight initiation distances, as well as the distance fled. We found that birds show increasingly stronger antipredator behaviors as they initiate and increase their investment into reproduction, from non-reproductive stages to courting and brooding small, thermo-dependent chicks. However, once offspring gained thermal independence and freedom of movement, parents reduced their antipredator behaviors. Late breeders were more likely to flee from the approaching threat than early breeders. Altogether, our results demonstrate that parental antipredator responses are dynamic and shaped by the levels of investment into current reproduction, the ability of offspring to defend themselves, and the perceived future value of the brood.

## INTRODUCTION

When faced with approaching predators, prey must undertake decisions on whether to flee or not, and if they do, how close to allow predators to approach before initiating flight. Flight Initiation Distance (FID) is thus an estimation of the amount of risk a prey evaluates as acceptable (Ydenberg and Dill 1986). If the risk is perceived to be high, prey should initiate flight sooner, for example, when a predator displays higher intent in the approach, such as approaching quickly (Cooper 1997; Cooper et al. 2003; Stankowich and Blumstein 2005; Cooper and Whiting 2007; Bateman and Fleming 2011) or directly (Burger and Gochfeld 1990; Cooper 2003; Cooper et al. 2003; Smith-Castro and Rodewald 2010). However, it is not just the level of risk that determines when prey will initiate flight. Optimal escape distances are shaped by a trade-off between the fitness gains acquired from flight behavior vs. the fitness losses incurred by abandoning the current activity in which the individual is engaged at the time of predator disturbance (Cooper and Frederick 2007). For instance, fleeing an approaching predator can directly improve individual survival or limit injuries in prey-predator conflicts, but interferes with breeding activities such as territory establishment and defense (Jakobsson et al. 1995; Dunn et al. 2004; Møller et al. 2008; Hess et al. 2016), pairing and mate choice (Cooper and Wilson 2007; Cooper 2009; Ventura et al. 2021), nest building (Lima 2009), or offspring care (Montgomerie and Weatherhead 1988; Carter et al. 2009; Arroyo et al. 2017; Novčić and Parača 2022).

For breeding adults, the decision to stay and protect their young against a predator, potentially risking serious injuries or death, or to flee and abandon reproduction altogether, is conditioned by a trade-off between the value of current reproduction vs. that of future breeding opportunities (Williams 1966; Trivers 1972; Montgomerie and Weatherhead 1988). Long-lived species, such as seabirds, should generally favor adult survival over reproduction given their potential for reproducing in subsequent years. Yet, even in long-lived species, parental responses to approaching predators may be graded depending on the current value of the offspring. The "brood value hypothesis" predicts that parental commitment to offspring should increase (and thus FID decrease) at least up to a point, as the relative reproductive value of offspring increases with their higher likelihood of surviving to independence (Barash 1975; Andersson et al. 1980; Ackerman and Eadie 2003; Redmond et al. 2009; Svagelj et al. 2012). In addition, in many species, the ability to reproduce successfully a second time within a breeding season declines as the breeding season advances due a drastic reduction in the chance of successfully raising a brood to fledging, and so the cost of losing offspring to predation increases (Andersson et al. 1980; Montgomerie and Weatherhead 1988; Redmond et al. 2009). Accordingly, in birds, studies have suggested that in species with high re-nesting potential, parents should commit less effort to nest defense and have longer FIDs than those with low re-nesting potential (Dawkins and Carlisle 1976; Boucher 1977; Weatherhead 1979; Klvaňová et al. 2011; Svagelj et al. 2012). Alternatively, in many other species, the likelihood for parents to fledge or wean their offspring decreases considerably as the breeding season advances and resources become less abundant. Consequently, in these species where resource abundance is highly seasonally, late breeding parents should be less prone to defend their brood against predators. Thus, offspring of higher reproductive value should be those that are older, those that are impossible to replace due to the advancement of the breeding season, or those that, independently of their age, are more likely to reach maturity due to the advancement of the breeding season (Barash 1975; Andersson et al. 1980; Kleindorfer et al. 1996; Tryjanowski and Goławski 2004).

Beside offspring age and the advancement of the breeding season, parental investment into offspring defense is also known to peak depending on the intersection of highest brood vulnerability and highest brood value (which occurs at hatching/birth or at fledging/weaning, depending on the species). The dependence of offspring upon their parents for protection against predators (contingent on whether offspring are both altricial or precocial) appears to play an important role in shaping parental responses to predators (Barash 1975; Andersson et al. 1980). In species with precocial young, parental defense of offspring is usually highest immediately after hatching/birth and declines quickly as offspring near fledging/weaning and gain the physical independence necessary to flee predators on their own (killdeer, *Charadrius vociferus*, Brunton 1990; willow ptarmigan, *Lagopus lagopus*, Sandercock 1994; Savanna nightjar, *Caprimulgus affinis*, Tseng et al. 2017). In species with altricial offspring, parental investment in offspring defense continues to increase with offspring age, peaking later in the nestling phase prior to fledging/weaning when offspring begin to develop some independence in movement (fieldfare, *Turdus pilaris*, Andersson et al. 1980; Adélie Penguins, *Pugoscelis adeliae*, Wilson et al. 1991; meadow pipit, *Anthus pratensis*, Pavel 2006; African penguin, *Spheniscus demersus*, Pichegru et al. 2016). In bird species with altricial young, interruptions of parental care shortly after hatching (when altricial chicks heavily rely on parents for brooding), may also greatly increase the risk of nest failure. Parents of altricial chicks may thus be less willing to flee the nest upon the approach of a predator than in species with precocial young, whose offspring are frequently observed scattering upon the approach of a predator and might survive independently of the parent for a period of time (Andersson et al. 1980; Buitron 1983; Sandercock 1994).

Here, using non-lethal human approaches to mimic predation threat (Frid and Dill 2002; Beale and Monaghan 2004, see Hammer et al. 2022 for an example on king penguins), we tested, in a semi-altricial species, the king penguin (*Aptenodytes patagonicus*), how adult antipredator responses varied according to life history stage (molting, courting, incubating, brooding small vs.

large chicks) and associated investment into reproduction, the dependence of their offspring for protection, and the timing of reproduction (early vs. late breeders). King penguins are an interesting model to compare the fitness benefits vs. costs of flight between various life history stages, and more specifically investigate how flight responses or nest defense are shaped by the reproductive value of the brood. Successfully breeding king penguins only produce one egg every 14-16 months (i.e., the time needed to form a pair, lay a single egg, incubate, fledge the chick, and molt before subsequent reproduction) (Stonehouse 1960; Weimerskirch et al. 1992), and therefore the investment in reproduction, from egg incubation to chick fledging, is extremely high. Only adults that breed in the first half of the reproductive season (*i.e.*, early breeders) are usually successful at fledging a chick (Weimerskirch et al. 1992), since late breeders are constrained by the arriving winter and their breeding success is virtually null (Van Heezik et al. 1994; Olsson 1996). Offspring of late breeders often have too few body reserves to survival the austral winter until their parents start feeding them again after the winter (Geiger et al. 2012, Stier et. 2014). Thus, re-nesting potential is extremely low in this species. Therefore, the relative value of the brood is expected to be higher for early breeding birds (Barash 1975; Andersson et al. 1980; Montgomerie and Weatherhead 1988; Redmond et al. 2009). The chicks are hatched in a semi-altricial state and remain highly dependent on their parents for feeding, protection, and body heat for the first thirty days. After some thirty days they are able to achieve thermal independence when they grow in a coat of down (Stonehouse 1960). At this age, the chicks become less reliant on parental care as they can cluster into "creches", decreasing their risk of predation without necessary protection of their parents (Stonehouse 1960; Le Bohec et al. 2005). Offspring stay in creches for several months during winter until their parents start feeding them again at the end of the austral winter and they finish growing (Cherel et al. 1985; Weimerskirch et al. 1992: Stier et al. 2014).

Breeding king penguins are also highly aggressive, and territorial aggression increases upon the hatching of a chick, reflecting increased brood defense (Côté 2000).

We approached >500 individual penguins at different life history stages and measured their behavioral reaction to our approach, including the distances at which they began to show signs of alertness, i.e. head-turn towards the experimenter (Alert Distance, AD), the distance at which they initiated flight (FID), and the distance over which they fled once flight occurred (Distance Fled, DF). We also recorded the number of aggressive interactions emitted towards the approaching experimenter. According to life history theory on the tradeoff between current and residual fitness (Roff 1992; Stearns 1992), we predicted that (i) breeders should be less prone than non-breeders to flee from approaching predators; (ii) among breeders, parents should invest more into offspring defense and be less prone to flee as offspring age (brood value hypothesis), at least until the offspring can also take part in anti-predator defenses (Le Bohec et al. 2005); (iii) for a similar investment into reproduction, breeders with lower prospective to fledge their young should invest less into offspring defense (i.e. late breeders vs. early breeders).

To test these predictions, we first compared bird behavioral reactions across a range of life history stages committed to reproductive activities (courting birds, birds in breeding pairs, incubating birds, and brooding birds), or not (molting birds, and birds not engaged in reproduction). We expected birds in breeding pairs to have shorter FID and DF and to be more aggressive than molting birds and birds not engaged in reproductive activities. We further expected birds with eggs and chicks to have the shortest FID and DF, to be less susceptible to initiate flight, and to be the most aggressive due to their increased reproductive investment.

We then compared the response of parents rearing chicks of different ages (less or more than 30 days old). We expected to find higher nest defense and shorter flight responses in birds caring for older chicks of higher reproductive value, and in early breeding birds for which the likelihood of successful reproduction is higher. Alternatively, increasing parental offspring defense and decreasing flight behavior from hatching to early brooding stages when chicks are the most vulnerable and reproductive value highest, followed by a subsequent decrease in chick defenses and increase in flight behavior once chicks gain physical autonomy (and are able to flee from predators independently) might indicate that king penguins conform to antipredator responses expected from altricial species.

Finally, we tested for differences in parental commitment to reproduction between early and late breeding adults, since in this species only early breeding adults (during the first half of the reproductive season) have a reasonable chance of successfully raising a chick (Weimerskirch et al. 1992; Van Heezik et al. 1994; Olsson 1996). We expect late breeders to be less willing to invest into offspring defense and show longer FID and DF, and increased susceptibility to initiate flight upon the approach of an experimenter.

## METHODS

### *Study site and animals*

This study was conducted during the 2010-2011 and 2011-2012 austral summers in the king penguin colony of "La Baie du Marin" (22 000 breeding pairs, Barbraud et al. 2020) located in the Crozet Archipelago (Possession Island, 46°25'S – 51°52'E).

In a first investigation of how birds reacted to an approaching threat at various life history stages and to further test the brood value hypothesis among breeding adults, we approached 498 adult king penguins (data set 1). Individual birds were either molting birds (n = 24), birds not engaged in reproduction (n = 26), courting birds (n = 20), birds settled on their final breeding territory with their partner (n = 17), incubating birds (n = 283), birds brooding small, thermally dependent chicks (TDC; <30-days old) (n = 78), or birds brooding older, thermally-independent chicks (TIC; >30-days old) (n = 50). TDC depend on their parents for thermoregulation and are kept warm against the adult in the brood pouch until they have grown a full set of down feathers and are able to be left alone for longer periods while the adults forage at sea. All breeding birds measured in this first analysis were early breeding birds.

In a second investigation, to test how breeding prospects affected antipredator behavioral reactions to an approaching threat, we compared the behavior of an additional set of birds that started incubating early in the season (November, N=50) to birds that started incubating late in the breeding season (March, N=50) (data set 2). Individuals were chosen at random: we had no information on their age and sex, nor the exact laying dates and shift for all incubating and brooding birds.

### *Approach protocol*

All approaches occurred between 8:00 am and 6:00 pm and were carried out by the same observer (B.G.) wearing the same clothing for each approach. Approaches were only started in clear sight of the target animal when the animal was at rest (not engaged in aggressive, grooming, sleep, courting, or offspring care). All aggression directed towards the approaching experimenter during the approach was recorded in order to measure territoriality and brood defense (see below). All occurrences of threats and physical blows towards the experimenter were counted. Aggression during approach for birds in breeding pairs was not recorded. Distances between the observer and the focal bird were measured using a laser telemeter (Leica DISTO™ D5 Lasermeter, Leica Geosystems AG, Hexagon, Sweden, ± 1mm accuracy, 0.1mm resolution over 0.05-120m range). The vast majority of the approaches (488/598, 82%) were standardized to an 18-m starting distance to control for the commonly observed effect of starting distance on AD and FID (Dumont et al. 2015; Blumstein, 2010; Blumstein, 2003). However, not all starting distances could be standardized due to the terrain or location of the bird (see also Samia et al. 2017). Courting birds and birds in breeding pairs were approached between 12 m and 20 m (mean 14.3 m and 13.8 m, respectively), and some incubators between 11 m and 20 m (mean 17.8 m). The starting distance for brooders of TDC varied between 16 m and 18 m (mean 17.8 m). Because a preliminary analysis showed no strong correlation between starting distance and AD or FID within any of the breeding stages across data sets (AD: courting birds, $r_{(df=18)} = 0.25$, $P = 0.29$, birds in breeding pairs, $r_{15} = 0.11$, $P = 0.67$, incubating birds, $r_{381} = 0.07$, $P = 0.19$, brooders of TDC, $r_{76} = 0.11$, $P = 0.35$, brooders of TIC, $r_{48} = 0.29$, $P = 0.04$; FID: courting birds, $r_{18} = 0.11$, $P = 0.65$, birds in settled pairs, $r_{15} = -0.23$, $P = 0.37$, incubating birds, $r_{381} = -0.04$, $P = 0.40$, brooders of TDC, $r_{76} = -0.07$, $P = 0.50$, brooders of TIC, $r_{48} = 0.01$, $P = 0.91$), we did not control for starting distance in later analyses.

Approaches followed a direct trajectory, with the observer walking at a regular speed until the occurrence of AD (observable due to movement of the focal animal's head in the direction of the experimenter; Fernández-Juricic et al. 2001), and FID (when the target animal initiated an attempt to flee). At these two times, the experimenter took a standardized one-minute pause to record information. Animals were approached until close contact. If flight occurred after close contact, the distance over which the focal animal fled (DF) was recorded. Because birds did not systematically flee upon approach until close contact (FID=0), FID was considered in two different sets of analyses. The first explored the decision to initiate flight as a binary variable (0/1; no flight/flight), and the second explored FID in those animals that choose to initiate flight.

We calculated speed of approach in three separate periods: the speed of approach prior to AD (mean ± SE; 0.57 ±0.00 m/s, range = 0.27 – 1.23 m/s), the speed of approach prior to FID (0.48 ± 0.01 m/s, range = 0.03 – 1.06 m/s), and the speed of approach prior to contact (0.46 ± 0.01 m/s, range = 0.13 – 0.83 m/s). In addition, we systematically recorded weather conditions before each approach to control to control for potential effects of climate conditions on bird behavior (Couchoux and Cresswell 2012). Air temperature was measured to the nearest 1°C and the levels of wind and rain were scored between 0 (none) and 2 (strong or heavy) with intermediate levels (0.5, 1, 1.5) being permitted.

***Statistical analyses***

We used separate mixed effects models with appropriate error distributions (Linear Mixed Models, LMM, or Generalized Linear Mixed Models, GLMM; package "lme4' in R Bates et al. 2015) to test how AD, FID, the decision to flee (binomial, 0/1), DF and bird aggression towards the experimenter (0/1) were shaped by (1) life history stages (molting, birds not engaged in reproductive activities, courting birds, birds in breeding pairs, incubating birds, birds brooding TDC, and birds brooding TIC) (data set 1); and (2) reproductive timing (early vs. late incubators) (data set 2). A final model tested the relationship between aggression emitted by the focal bird towards the approaching experimenter and the decision to flee (data set 1). Because a previous study found that physical blows were 3.2 times more costly in terms of energy expenditure than threat behavior (Viera et al. 2011), we multiplied physical blows (flipper blows and pecking) by 3.2 in order to weigh these behaviors more heavily than threat behaviors (gapping and bill pointing), thus better capturing parental commitment and energy expenditure into brood defense. We then tested whether weighted aggression affected the probability to flee (0/1) using a GLMM. Because we expected this might change between breeding stages due to different parental commitment to the offspring, we added the interaction between aggression and breeding stage in the model.

In general, because AD, FID, FD, and aggressive behaviors can be influenced by external factors such as time of day, time of day$^2$ (predators are often more active at dawn and dusk, leading to stronger responses in the target prey; Petelle et al. 2013; Piratelli et al. 2015; Ferguson et al. 2019), weather (rain, wind, and temperature) conditions (Couchoux and Cresswell 2012; Hammer et al. 2022, 2023), and approach speed, we controlled for these potential effects in all analyses and present full models with standardized effect sizes (Z-scores) in our results. In all models, bird location in the colony was included as random factor to control for potential similarities in behavioral responses of individuals being measured in the same areas, explained for example by a landscape of fear, with some areas being more exposed to predators or human disturbances. We removed colony location from models which did not converge due to low amounts of variation explained by this factor, resorting to simpler (generalized) linear models (LM or GLM). For 21 birds, colony location was not recorded and these were thus dropped from analyses where colony location was specified.

All statistical analyses were performed in R 4.1.2 (R Development Core Team 2021). Results are presented as means ± SE. Models were run using the *lmer* and *glmer* functions (package 'lme4' v.1.-27.1; Bates et al. 2015). All continuous covariates were scaled and centered prior to inclusion in the models to facilitate their interpretation (Schielzeth, 2010). In models with a gaussian distribution, the significance of fixed effects was tested with the function *anova* from the package 'lmerTest' (v.3.1-3; Kuznetsova et al. 2017), which utilizes F tests with Satterthwaite estimation for the denominator degree of freedom. For models with a binomial distribution, the function *Anova* was used, which utilizes Wald chi-square tests to test the significance of fixed effects. The package emmeans (v.1.10.5; Lenth 2021) was used for post-hoc comparisons using the Kenward-Roger degree of freedom method and Tukey method for p-value adjustment. Variance Inflation Factors (VIF) were used to inspect the collinearity between all independent variables, with a cut-off of less than 3 (Zuur et al. 2010). These analyses showed that, in the early/late breeder models, temperature was collinear with reproductive stage (weather was warmer later in the season when late breeders were initiating breeding), so we excluded temperature from those models.

Models' residuals were inspected visually for normality using the 'fitdist' function of the 'fitditrplus' package in R (v.1.2-1; Delignette-Muller and Dutang, 2015). FID and FD values were log-transformed before analyses, as this led to a better distribution of model residuals. Binomial data was analyzed in generalized linear models specifying a binomial error distribution (logit link). Although we approached a total of 498 individual birds for comparison across the life history stages (data set 1), and an additional 100 individuals to compare early and late breeders (data set 2), sample sizes for each model vary due to a lack of complete information for some of the variables depending on analyses. For example, distances fled (FD) or locations were not systematically recorded, so that samples sizes vary across models. Sample sizes are thus reported for each model in the results. The 'sjPlot' v2.8.16 package (Lüdecke 2024) was used to estimate marginal $R^2$ (fixed effects) and conditional $R^2$ (fixed and random effects) coefficients for mixed effects models, adjusted-$R^2$ coefficients ($R^2_{adj}$) for simple linear models, and Tjur's discrimination coefficients (Tjur's $R^2$) for binary outcomes (Tjur 2009). Full model results are provided as online supplementary materials.

### *Ethics statement*

No animal was caught or manipulated over the course of this study. The research was approved by the Ethical Committee of the Institut Polaire Français – Paul-Emile Victor. Authorizations to enter the colony and approach birds were obtained from Terres Australes et Antarctiques Françaises. The observations complied with the current laws of France. No eggs or chicks were abandoned during the course of this study.

## RESULTS

### *Comparing life history stages*

Over all life history stages, individuals became alert on average at a distance of 7.27 ± 0.09 m (mean ± SE, range = 2.18 – 14.7 m, N = 498). Controlling for other variables in the model (Fig. 1, OSM 1), there were significant differences in AD among the life history stages (LMM, $F_{6,432.69}$ = 4.92, $P$<0.0001, N = 450, Marginal $R^2$ / Conditional $R^2$ = 0.11 / 0.18). In general, molting birds and non-reproductive birds had higher AD than reproductive birds (Fig. 1). Molting and non-breeding birds did not differ significantly in AD, nor did non-breeding and courting birds, or courting birds, birds in settled pairs, and incubators and brooders (Fig. 1). There was a slight negative effect of approach speed between starting distance and AD (estimate ± SE = -0.20 ± 0.09; $F_{1,433.8}$ = 5.23, $P$ < 0.0001), but no significant effect of weather or time of day on AD (Fig 1).

Incubators initiated flight in 45.2% of approaches (128/283), brooders of TDC in 52.6% of approaches (41/78), brooders of TIC in 94% of approaches (47/50), and all other life history stages in 100% of approaches (24/24 molting birds, 26/26 birds not engaged in reproductive activities, 20/20 courting birds, and 17/17 birds in breeding pairs). Focusing on groups for which flight initiation was not 100% and controlling for other variables in the model, there were significant differences between the life history stages (GLM, $X^2$ = 45.2, $P$<0.0001, N = 393, OSM 2). Incubators and brooders of TDC did not differ in their probability to initiate flight (Tukey HSD, Z = -1.72, $P$ = 0.199), but both fled significantly less often than brooders of TIC (incubator to brooders of TIC: Tukey HSD, $Z$ = -4.90, $P$<0.0001; brooders of TDC to brooders of TIC: Tukey HSD, $Z$ = -4.01, $P$<0.001). Here again, there was a negative effect of approach speed between AD and FID (odds ratio = 0.67, CI [0.51-0.87]; $X^2$ = 9.0, $P$ = 0.003, N = 393), but no other significant effect of other covariables on FID (Fig 2).

For those birds that did initiate flight (N = 303), on average and over all life history stages, flight was initiated at a distance of 3.10 ± (SE) 0.13 m (range = 0.2 – 14.7 m). Controlling for other variables in the model, there were significant differences in FID between the life history stages (LMM, $F_{6,239.9}$ = 13.3, $P$<0.0001, N = 260, Marginal $R^2$ / Conditional $R^2$ = 0.33 / 0.35, OSM 3). On average, molting bird and birds not engaged in reproductive activities had higher (and similar) FIDs than birds engaged in breeding activities (Fig 3). Overall, FID decreased progressively with advancing breeding stage, not being significantly different between stages from courtship to chick-brooding. Interestingly, FID tended to increase again in brooders of TDC (+34%) and TIC (+32%) compared to incubators, though not significantly (P = 0.10 and 0.06, respectively) (Fig 3).

When flight distance could be recorded (N = 201 cases), birds fled over an average distance of 1.9 ± (SE) 0.20 m (range = 0.19 – 15.0 m) from the experimenter after contact. Distance Fled (DF) varied significantly among the life history stages (LMM, $F_{5,93.44} = 96.6$, $P<0.0001$, N = 140, Marginal $R^2$ / Conditional $R^2 = 0.85$ / 0.85, OSM 4). In general, molting birds and birds not engaged in reproductive activities had significantly higher DF than reproductive birds (Fig. 4). Courting birds and brooders of TIC did not differ significantly in terms of DF, which were all significantly higher than incubators and brooders of TDC (Fig. 4).

For approaches where aggression was recorded, and controlling for other variables in the model, not all life history stages behaved aggressively towards the experimenter (GLM; $X^2 = 172.9$, $P<0.0001$, N = 454, $R^2 = 0.41$, OSM 5). Molting birds were never aggressive (0/24), birds not in engaged in reproductive activities were aggressive in 4% of approaches (1/26), 5% for courting birds (1/20), and 18% in brooders of TIC (9/50). In contrast, incubating birds and birds brooding TDC were the most aggressive, with aggression occurring in 80% (219/282) and 79% (61/77) of approaches, respectively. The probability of being aggressive was negatively related to the initial speed of the approach (between starting distance and AD; odds ratio = 0.77, CI [0.61-0.98]; $X^2 = 4.368$, $P<0.037$) and birds tended to be less aggressive in sunnier conditions (odds ratio = 0.77, CI [0.59-1.01]; $X^2 = 4.368$, $P=0.06$).

### *Comparing early and late incubating breeders*

Controlling for other effects in the model, AD was not significantly different between early and late breeders (LMM, $F_{1,88.54} = 1.79$, $P = 0.185$, N = 98, Marginal $R^2$ / Conditional $R^2 = 0.19$ / 0.23, OSM 6). Early breeders alerted at 6.61 ± 0.23 m (range = 3.72 – 11.2 m) on average, whereas late breeders alerted at 7.01 ± 0.24 m (range = 2.61 – 9.98 m) on average. In contrast, late breeding birds tended to be more likely to flee upon the approach of an experimenter (GLM, odds ratio =

3.45, CI [0.92-14.16]; $X^2$ = 3.38, $P$ = 0.07, N = 99, $R^2$ = 0.10, OSM 7), a difference which became significant when other non-significant effects (climate and hour variables) were removed from the model ($X^2$ = 5.85, $P$ = 0.02, $R^2$ = 0.08). In total, 52% (26/50) of late breeders fled when approached, whereas in comparison, only 44% (22/50) of early breeders fled when approached. Approach speed between AD and FID had a negative effect on the probability to flee (odds ratio = 0.23, CI [0.07-0.65]; $X^2$ = 7.94, $P$ = 0.005, N = 99).

For birds that did flee, controlling for other effects in the model, we found no significant different between early and late breeders in FID (LM; $F_{1,39}$ = 0.004, $P$ = 0.893, $N$ = 47, $R^2$ = 0.37, OSM 8). Approach speed had a positive effect on FID (estimate = 0.65 ± 0.16, $F_{1,39}$ = 27.64, $P$ < 0.0001). Early breeders initiated flight at an average of 2.05 ± 0.40 m (range = 0.67 – 8.46 m), and late breeders at an average of 2.62 ± 0.25 m (range = 1.01 – 6.76 m). Similarly, the distance fled (FD) did not differ significantly between early and late breeders (LM, $F_{1,37}$ = 059, $P$ = 0.449, N = 29, $R^2$ = 0.19, OSM 9), after controlling for other effects in the model. Note however, that the sample size for this model was low, and caution is required with interpretation. Early breeders fled 0.11 ± 0.02 m (range = 0.05 – 0.30 m) on average, and late breeders fled 0.14 ± 0.03 m (range = 0.05 – 0.50 m) on average.

We found no significant difference in the probability of early and late breeder being aggressive to the approaching experimenter (GLM, $X^2$ = 0.25, $P$ = 0.612, N = 100, $R^2$ = 0.17, OSM 10), after controlling for other effects in the model. The probability of aggression increased with the initial speed of the approach (between starting distance and AD; odds ratio = 4.79, CI [1.39-21.82]; $X^2$ = 6.44, $P$ = 0.011). Early breeders were aggressive towards the approaching experimenter in 84% of the approaches (42/50), and late breeders in 86% (43/50).

***Are fleeing and aggressive behaviors mutually exclusive traits?***

Focusing only on incubating and brooding birds for which more than one aggressive event was recorded per group, we found a significant interactive effect of weighted aggression and breeding stage on the birds' probability to flee (GLMM, $X^2$ = 7.19, $P$ = 0.027, N = 399, Marginal $R^2$ / Conditional $R^2$ = 0.35 / 0.36, OSM 11). Less aggressive birds more likely to flee the approaching experimenter, but this effect was increasingly marked as breeding stage advanced (incubating < brooding TDC < brooding TIC) (Fig. 4).

## DISCUSSION

In this study, we used non-lethal human approaches to mimic predation threat and test how antipredator behaviors varied across a range of life history stages in king penguins, some of which were committed to reproduction (courting birds, birds in breeding pairs, incubating birds, and chick-rearing birds) and others not (molting birds, and birds not engaged in reproduction). We found that birds are increasingly reluctant to initiate flight and flee shorter distances as reproduction advances from non-reproductive stages to courting and brooding small, thermo-dependent chicks. However, the probability to flee and flight initiation distance increases again once offspring have acquired thermo-independence and freedom of movement. Late breeders tended to be more likely to flee than early breeders, and flight negatively correlated with bird aggressiveness across incubating and brooding stages.

### *Comparing antipredator behaviors across life history stages*

Life history stage significantly affected bird antipredator behavior including AD, FID, the decision to flee or not, and the probability to emit aggression towards the approaching threat. This would be expected if the different life history stages varied in their costs and benefits of flight (Ydenberg and Dill 1986; Cooper and Frederick 2007). Molting birds and birds not actively engaged in reproductive activities had the longest FD, and the longest FID with courting birds. They also systematically initiated flight upon approach and displayed no aggressive behavior towards the experimenter. Birds in these life history stages had the highest fitness benefit/cost ratio of flight, as they were not defending a territory or courting potential mates, and so there was no trade-off between reproduction and individual survival. In contrast, birds engaged in reproduction appeared reluctant to flee, with the lowest AD and fleeing probabilities generally found in incubating birds and birds brooding small chicks. These birds face a trade-off between the fitness gains from reproduction vs. the fitness costs suffered from potential injury. By fleeing, they risk losing their reproductive investment compared to birds having not yet heavily invested into reproduction. It is somewhat surprising that courting birds, for which the investment in reproduction was relatively low showed overall similar AD and FID levels than later reproductive stages of higher reproductive investment (*e.g.*, chick-brooders). One explanation may be that courting birds are reluctant to abandon a tentative mate, for which a complex and ritualized process of mate choice occurs, suggesting that the process of mate choice may be a significant reproductive investment in itself. Indeed, in king penguin, mate choice involves a complex ritualized display over several days (Stonehouse 1960), and the careful evaluation of ornaments (bright-yellow and UV beak spots, orange to rusty brown feather patches and breast patch) known to honestly signal mate quality (Viblanc et al. 2016; Schull et al. 2016; Jouventin & Dobson 2017; Schull et al. 2018).

The cost of losing a partner to fleeing due to intra-sexual mate competition in primarily male-biased colonies (including our study colony; Descamps et al 2009; Bordier et al. 2013) may be particularly exacerbated in males (Keddar et al. 2013).

Taken at face value, selection for predator-response decisions based on past investments may seem like a Concorde fallacy (Dawkins & Carlisle 1976). However, in king penguins, this might be explained by the fact that on-land predation is seldom lethal for healthy adults, and mostly concerns targets chicks and eggs (Descamps et al. 2005). Thus, the costs of potential injuries may outweigh that of losing the egg or offspring for a species where reproduction is notoriously complex and requires bi-parental commitment for over a year to succeed (Stonehouse 1960). For incubating birds and brooders of TDC, distances fled (DFs) were substantially shorter than for other life history stages. This is because, in king penguin, incubating birds and brooders of TDC necessarily flee with the egg/young chick on top of their feet, restricting their movement. The most they can do without completely abandoning the offspring is to waddle a couple of meters and cluster with neighboring conspecifics within the colony, benefitting from predation dilution, confusion, mobbing, and selfish herd effects (Hamilton 1971; Cresswell 1994; Boland 2003; Dias 2006; Quinn and Cresswell 2006; Graw and Manser 2007; Olson et al. 2013; Hammer et al. 2023). In contrast, other life history stages not constrained in their movements were able to flee over greater distances. Note that the anti-predator benefits of fleeing might be substantially different in other ground-nesting birds during breeding. For instance, in waders, birds are generally more risk-averse and likely to flee during breeding than during other life history stages (Mikula et al. 2018a,b). Fleeing from an approaching predator in waders may actually improve breeding success, as it minimizes nest-detection probability by predators (Lima 2009). In contrast, fleeing from a predator in king penguins means abandoning the egg or chick (kept on top of the adult's feet) in the open that equals to breeding failure.

In king penguin, it is clear that incubators and brooders of TDC face the strongest fitness trade-off, as they could potentially abandon reproduction altogether and flee over greater distances; however, this strategy was never found to occur. Two, non-mutually exclusive, explanation might explain this result. First, the actual fitness benefits of fleeing from predation in a king penguin colony are lower than the fitness benefits of clustering with other conspecifics. An isolated adult may indeed be at more risk of being predated than if it clusters with aggressive neighbors (Côté 2000), mobbing and overwhelming the predator. A previous study for instance found that central areas of higher bird density were less likely to be predated by giant petrels than sparser peripheral areas or beaches (Descamps et al. 2005), where these predators may more easily isolate individual penguins (TLH, PB, BG, JPR, RG, VAV; personal observations). Second, as mentioned above, the fitness costs of abandoning current reproduction may actually be high in this species, despite it being long-lived. Not only is the energy commitment to reproduction is extremely high (Cherel et al. 1988; Groscolas et al. 2000, 2008; Groscolas and Robin 2001), successful reproduction only occurs every second to third year in this species (Weimerskirch et al. 1992; Van Heezik et al. 1994; Olsson 1996). Without parental care, eggs and thermally dependent king penguin chicks are highly vulnerable to cold and to predation (Hunter 1991; Descamps et al. 2005). Accordingly, incubators and brooders of TDC were also the only life history stages that did not flee in approximately half the approaches. Rather, these birds adopted an aggressive stance towards the approaching experimenter, defending their egg/young chick and territory position. Similar results on parental commitment were found in Magellanic (*Spheniscus magellanicus*) (Villanueva et al. 2014) and Adélie (*Pygoscelis adeliae*) penguins (Wilson et al. 1991).

AD was higher in molting birds and birds not actively engaged in reproductive activities, than in all other reproductive stages (courting birds, birds in breeding pairs just settled onto their territory, incubating birds, and brooding birds). This result is somewhat surprising given that studies generally show that animals increase vigilance behavior when breeding. For example, reproductive females with accompanying offspring often display increased vigilance behavior (e.g., in a range of African mammals, Burger and Gochfeld 1994; elk, *Cervus elaphus*, Childress and Lung 2003; and eastern grey kangaroo, *Macropus giganteus*, Carter et al. 2009). Similarly, incubating Magellanic penguins show higher AD than molting and settling birds (Villanueva et al. 2014). One explanation may be the investment into reproduction (including courtship, egg and chick care) is done at the detriment of time spent in vigilance behaviors, explaining shorter AD. The distracted prey hypothesis (Chan et al. 2010; Petrelli et al. 2017) proposes that the processing of social cues and interactions with conspecifics may interfere with predator vigilance (Mooring and Hart 1995; Yee et al. 2013), and is indeed found to partly shape vigilance behavior in king penguins (Hammer et al. 2023). In contrast, molting birds, and non-reproductive birds are not engaged in acquiring a mate or territory, and may more devote a larger part of their time budget to predator vigilance. An alternative explanation may be related to information transmission: molting and non-breeding birds typically form mobile groups on the outskirts of the breeding colony. This dynamic social environment is likely to transmit information on predator presence to the surrounding neighbors rapidly (many-eyes hypothesis; Pulliam 1973; Lima and Dill 1990; Burger and Gochfeld 1991; Mayer et al. 2019), since predators entering a grouping of molting or non-breeding penguins will create a movement ripple travelling through the group (TLH, PB, JPR, BG, VAV; personal observations).

*Antipredator responses and brood value*

Our results show that (1) incubating birds had the lowest FID (lower than brooders of TDC and TIC), (2) FD was shorter in incubating birds and brooders of TDC than brooders of TIC, (3) incubating birds and brooders of TDC defended the brood with equal presence of aggression upon approach by the experimenter (~50% not fleeing, and ~80% displaying aggression), and (4) brooders of TIC were less likely to stay and defend their young aggressively against the approaching experimenter (only 6% not fleeing, and 18% displaying aggression). In addition, brooders of TIC mostly fled independently of their chicks, while brooders of TDC always fled with their chick on top of their feet, resulting in much shorter FD. These results agree with the "brood value hypothesis", which predicts low FID and FD and higher rates of brood defense when brood value is high (Trivers 1972; Barash 1975; Montgomerie and Weatherhead 1988; Frid and Dill 2002). In line with antipredator behaviors, territorial brood defense is known to increase from incubation to young, non-thermally emancipated chick brooding in king penguins (Côté 2000), showing an increase in brood value as the brood ages.

In species with altricial young, investment into offspring defense increases gradually from incubation to brooding (Barash 1975), with a peak at the end of the nestling period, before sharply decreasing when the offspring can flee from predators themselves (Andersson et al. 1980; Redmond et al. 2009; Strnadová et al. 2018). Our results are consistent with those found in other species including other penguin species (Wilson et al. 1991; Cevasco et al. 2001; Villanueva et al. 2014; Pichegru et al. 2016), mammals (Koskela et al. 2000), and fish (Thünken et al. 2010). Mechanistically, changes in parental commitment to offspring have been suggested to be underpinned by attenuated parental stress responses to external threats when the relative reproductive value of offspring is high (Lendvai et al. 2007; Lendvai and Chastel 2008; Bókony et al. 2009; Schmid et al. 2013; Viblanc et al. 2015). This appears to be the case in king penguin where hormonal (Viblanc et al. 2016) and heart rate (Viblanc et al. 2015) stress responses to non-lethal human approaches are attenuated in brooding vs. incubating birds, indeed suggesting a redirection of parental physiology geared towards offspring care rather than adult survival.

***Comparing early and late breeders***

Higher fitness by individuals initiating their reproduction earlier in the season is frequently observed in birds and mammals (Bednekoff 1996; Kokko 1999; Götmark 2002; Williams et al. 2014; Germain et al. 2015; de Villemereuil et al. 2020), and early breeders are often found to defend their brood more than late breeders, for example in red-winged blackbirds *Agelaius phoeniceus* (Biermann and Robertson 1981), house sparrows *Passer domesticus* (Klvaňová et al. 2011), eastern kingbirds *Tyrannus tyrannus* (Redmond et al. 2009), and great tits *Parus major* (Onnebrink and Curio 1991, Redmond et al. 2009; Klvaňová et al. 2011). We expected late breeding king penguins to show lower commitment to their brood than early breeders given their extremely low chances of successful reproduction (Weimerskirch et al. 1992; Van Heezik et al. 1994; Olsson 1996). This expectation was only partially borne out by our results. While AD, FID, FD, and aggression were comparable between early and late breeders, late breeders were indeed more likely to initiate flight when approached than early breeders. This is consistent with results from other seabirds, such as snow petrels (*Pagodroma nivea*) (Goutte et al. 2011). The fact that only the probability to flee seemed to be affected by seasonal timing underlines the usefulness of measuring antipredator responses to an approaching experiment using a gradual scale and distances. It suggests that the response of last resort might be the most important to gain insight on the trade-off between investment into reproduction and survival, as demonstrated here in incubating king penguins.

### *Conclusions*

Taken together our results show that antipredator behaviors changed in concert with changes the cost of escape, in accordance with optimal escape decision. Non-reproductive birds were more prone to flee, fled further, and were less aggressive in response to an approaching experimenter. Reproductive birds, in particular incubating and brooding birds of young chicks were the least prone to flee, and frequently stayed and exhibited aggressive defensive behavior. Breeding birds with higher brood value invested more in offspring defense and were less likely to flee from an approaching experimenter, yet offspring defense reduced as chicks gained independence of movement, in accordance with results seen in species with altricial young. Finally, late breeding birds, with much lower reproductive success than early breeding birds were more likely to flee upon approach, suggesting lower perceived brood value. Our results demonstrate that antipredator behavioral responses are dynamic and shaped by the costs of escape, in particular by current investment into reproduction and the perceived brood value. Further research is needed to understand the observed differences between breeding stages, notably between courting and later reproductive stages. Evaluating the potential cost of losing a tentative mate due to flight by monitoring birds for re-pairing potential with the same or a different mate may prove particularly insightful in determining the commitment invested into reproduction at early stages of the breeding cycle.

## ACKNOWLEDGMENTS

This work was financially supported by the French Polar Institute (IPEV, program 119 ECONERGY), and by the French National Center for Scientific Research (CNRS). We are especially grateful to all field assistants and personnel for their help in the field. IPEV and the Terres Australes et Antarctiques Françaises (TAAF) provided unparalleled logistic support in the field, and we thank them thoroughly for their efforts in upholding polar research. We thank two anonymous reviewers for constructive comments on a previous version of this paper. THL was supported by a PhD scholarship from the initiative d'excellence (IdEX), investissements d'avenir, ministère de l'enseignement supérieur, de la recherche et de l'innovation.

## LITERATURE CITED

Ackerman JT, Eadie JMA (2003) Current versus future reproduction: an experimental test of parental investment decisions using nest desertion by mallards (*Anas platyrhynchos*). Behav Ecol Sociobiol 54:264–273. https://doi.org/10.1007/s00265-003-0628-x

Andersson M, Wiklund CG, Rundgren H (1980) Parental defence of offspring: a model and an example. Anim Behav 28:536–542

Arroyo B, Mougeot F, Bretagnolle V (2017) Individual variation in behavioural responsiveness to humans leads to differences in breeding success and long-term population phenotypic changes. Ecol Lett 20:317–325. https://doi.org/10.1111/ele.12729

Barash DP (1975) Evolutionary aspects of parental behaviour: distraction behaviour of the Alpine Accentor. Wilson Bull 87:367–373

Barbraud C, Delord K, Bost CA, et al (2020) Population trends of penguins in the French Southern Territories. Polar Biol 43:835–850. https://doi.org/10.1007/s00300-020-02691-6

Bateman PW, Fleming PA (2011) Who are you looking at? Hadeda ibises use direction of gaze, head orientation and approach speed in their risk assessment of a potential predator. J Zool 285:316–323. https://doi.org/10.1111/j.1469-7998.2011.00846.x

Bates D, Mächler M, Bolker BM, Walker SC (2015) Fitting linear mixed-effects models using lme4. J Stat Softw 67:1–48. https://doi.org/10.18637/jss.v067.i01

Beale CM, Monaghan P (2004) Human disturbance: people as predation-free predators? J Appl Ecol 41:335–343. https://doi.org/10.1111/j.0021-8901.2004.00900.x

Bednekoff PA (1996) Risk-sensitive foraging, fitness, and life histories: where does reproduction fit into the big picture? Am Zool 36:471–483. https://doi.org/10.1093/icb/36.4.471

Biermann GC, Robertson RJ (1981) An increase in parental investment during the breeding season. Anim Behav 29:487–489. https://doi.org/10.1016/S0003-3472(81)80109-1

Bókony V, Lendvai ÁZ, Likér A, et al (2009) Stress response and the value of reproduction: are birds prudent parents? Am Nat 173:589–598. https://doi.org/10.1086/597610

Boland CRJ (2003) An experimental test of predator detection rates using groups of free-living emus. Ethology 109:209–222

Boucher DH (1977) On wasting parental investment. Am Nat 111:786–788. https://doi.org/10.1086/283207

Bordier C, Saraux C, Viblanc VA et al (2014) Inter-annual variability of fledgling sex ratio in king penguins. PLoS One 9: e114052.

Brunton DH (1990) The effects of nesting stage, sex, and type of predator on parental defense by killdeer (*Charadrius vociferous*): testing models of avian parental defense. Behav Ecol Sociobiol 26:181–190. https://doi.org/10.1007/BF00172085

Buitron D (1983) Variability in the responses of black-billed magpies to natural predators. Behaviour 87:209–236. https://doi.org/10.1163/156853983x00435

Burger J, Gochfeld M (1990) Risk discrimination of direct versus tangential approach by basking black iguanas (*Ctenosaura similis*): variation as a function of human exposure. J Comp Psychol 104:388–394. https://doi.org/10.1037/0735-7036.104.4.388

Burger J, Gochfeld M (1994) Vigilance in African mammals: differences among mothers, other females, and males. Behaviour 131:153–169. https://doi.org/10.1163/156853994X00415

Burger J, Gochfeld M (1991) Human distance and birds: tolerance and response distances of resident and migrant species in India. Environ Conserv 18:158–165. https://doi.org/10.1017/S0376892900021743

Carter AJ, Pays O, Goldizen AW (2009) Individual variation in the relationship between vigilance and group size in eastern grey kangaroos. Behav Ecol Sociobiol 64:237–245. https://doi.org/10.1007/s00265-009-0840-4

Cevasco CM, Frere E, Gandini PE (2001) Intensidad de visitas como condicionante de la respuesta del pinguino de Magallanes (*Spheniscus magellanicus*) al disturbio humano. Ornitol Neotrop 12:75–81

Chan AAYH, Giraldo-Perez P, Smith S, Blumstein DT (2010) Anthropogenic noise affects risk assessment and attention: the distracted prey hypothesis. Biol Lett 6:458–461. https://doi.org/10.1098/rsbl.2009.1081

Cherel Y, Robin J, Walch O, et al (1988) Fasting in king penguin I . Hormonal and metabolic changes during breeding. Am J Physiol - Regul Integr Comp Physiol 254:R170–R177

Childress MJ, Lung MA (2003) Predation risk, gender and the group size effect: does elk vigilance depend upon the behaviour of conspecifics? Anim Behav 66:389–398. https://doi.org/10.1006/anbe.2003.2217

Cooper WE (2009) Flight initiation distance decreases during social activity in lizards (*Sceloporus virgatus*). Behav Ecol Sociobiol 63:1765–1771. https://doi.org/10.1007/s00265-009-0799-1

Cooper WE (1997) Factors affecting risk and cost of escape by the broad-headed skink (*Eumeces laticeps*): Predator speed, directness of approach, and female presence. Herpetologica 53:464–474

Cooper WE (2003) Risk factors affecting escape behavior by the desert iguana, *Dipsosaurus dorsalis*: Speed and directness of predator approach, degree of cover, direction of turning by a predator, and temperature. Can J Zool 81:979–984. https://doi.org/10.1139/z03-079

Cooper WE, Frederick WG (2007) Optimal flight initiation distance. J Theor Biol 244:59–67. https://doi.org/10.1016/j.jtbi.2006.07.011

Cooper WE, Pérez-Mellado V, Baird T, et al (2003) Effects of risk, cost, and their interaction on optimal escape by nonrefuging Bonaire whiptail lizards, *Cnemidophorus murinus*. Behav Ecol 14:288–293. https://doi.org/10.1093/beheco/14.2.288

Cooper WE, Whiting MJ (2007) Universal optimization of flight initiation distance and habitat-driven variation in escape tactics in a Namibian lizard assemblage. Ethology 113:661–672. https://doi.org/10.1111/j.1439-0310.2007.01363.x

Cooper WE, Wilson DS (2007) Sex and social costs of escaping in the striped plateau lizard *Sceloporus virgatus*. Behav Ecol 18:764–768. https://doi.org/10.1093/beheco/arm041

Côté SD (2000) Aggressiveness in king penguins in relation to reproductive status and territory location. Anim Behav 59:813–821. https://doi.org/10.1006/anbe.1999.1384

Couchoux C, Cresswell W (2012) Personality constraints versus flexible antipredation behaviors: How important is boldness in risk management of redshanks (*Tringa totanus*) foraging in a natural system? Behav Ecol 23:290–301. https://doi.org/10.1093/beheco/arr185

Cresswell W (1994) Flocking is an effective anti-predation strategy in redshanks, *Tringa totanus*. Anim Behav 47:433–442. https://doi.org/10.1006/anbe.1994.1057

Dawkins R, Carlisle TR (1976) Parental investment, mate desertion and a fallacy. Nature 262:131–133. https://doi.org/10.1038/262131a0

Descamps S, Le Bohec C, Le Maho Y, et al (2009) Relating demographic performance to breeding-site location in the King Penguin. The Condor 111: 81-87.

de Jong A, Magnhagen C, Thulin CG (2013) Variable flight initiation distance in incubating Eurasian curlew. Behav Ecol Sociobiol 67:1089–1096. https://doi.org/10.1007/s00265-013-1533-6

Dehn MM (1990) Vigilance for predators: detection and dilution effects. Behav Ecol Sociobiol 26:337–342

Descamps S, Gauthier-clerc M, Gendner J-P, Maho Y Le (2002) The annual breeding cycle of unbanded king penguins *Aptenodytes patagonicus* on Possession Island (Crozet). Avian Sci 2:87–98

Descamps S, Gauthier-Clerc M, Le Bohec C, et al (2005) Impact of predation on king penguin Aptenodytes patagonicus in Crozet Archipelago. Polar Biol 28:303–310. https://doi.org/10.1007/s00300-004-0684-3

de Villemereuil P, Charmantier A, Arlt D, Chevin L-M (2020) Fluctuating optimum and temporally variable selection on breeding date in birds and mammals. Proc Nat Acad Sci USA, 202009003. 10.1073/pnas.2009003117

Dias RI (2006) Effects of position and flock size on vigilance and foraging behaviour of the scaled dove *Columbina squammata*. Behav Processes 73:248–252. https://doi.org/10.1016/j.beproc.2006.06.002

Dunn M, Copelston M, Workman L (2004) Trade-offs and seasonal variation in territorial defence and predator evasion in the European Robin *Erithacus rubecula*. Ibis 146:77–84. https://doi.org/10.1111/j.1474-919X.2004.00221.x

Fernández-Juricic E, Jimenez MD, Lucas E (2001) Alert distance as an alternative measure of bird tolerance to human disturbance: implications for park design. Environ Conserv 28: 263-269.

Ferguson SM, Gilson LN, Bateman PW (2019) Look at the time: diel variation in the flight initiation distance of a nectarivorous bird. Behav Ecol Sociobiol 73:147

Frid A, Dill L (2002) Human-caused disturbance stimuli as a form of predation risk. Conserv Ecol 6:11. https://doi.org/10.1016/S0723-2020(86)80016-9

Geiger S, Le Vaillant M, Lebard T, Reichert S, Stier A, Le Maho Y, Criscuolo F (2012) Catching-up but telomere loss: half-opening the black box of growth and ageing trade-off in wild king penguin chicks. Mol Ecol 21: 1500-1510. https://doi.org/10.1111/j.1365-294X.2011.05331.x

Germain RR, Schuster R, Delmore KE, Arcese P (2015) Habitat preference facilitates successful early breeding in an open-cup nesting songbird. Funct Ecol 29:1522–1532. https://doi.org/10.1111/1365-2435.12461

Götmark F (2002) Predation by sparrowhawks favours early breeding and small broods in great tits. Oecologia 130:25–32. https://doi.org/10.1007/s004420100769

Goutte A, Antoine E, Chastel O (2011) Experimentally delayed hatching triggers a magnified stress response in a long-lived bird. Horm Behav 59:167–173. https://doi.org/10.1016/j.yhbeh.2010.11.004

Graw B, Manser MB (2007) The function of mobbing in cooperative meerkats. Anim Behav 74:507–517. https://doi.org/10.1016/j.anbehav.2006.11.021

Groscolas R, Decrock F, Thil MA, et al (2000) Refeeding signal in fasting-incubating king penguins: Changes in behavior and egg temperature. Am J Physiol - Regul Integr Comp Physiol 279:2104–2112. https://doi.org/10.1152/ajpregu.2000.279.6.r2104

Groscolas R, Lacroix A, Robin JP (2008) Spontaneous egg or chick abandonment in energy-depleted king penguins: A role for corticosterone and prolactin? Horm Behav 53:51–60. https://doi.org/10.1016/j.yhbeh.2007.08.010

Groscolas R, Robin J-P (2001) Long-term fasting and re-feeding in penguins. Comp Biochem Physiol Part A 128:645–655

Keddar I, Andris M, Bonadonna F & Dobson FS (2013) Male-biased mate competition in king penguin trio parades. Ethology 119: 389-396.

Hamilton WD (1971) Geometry for the selfish herd. J Theor Biol 31:295–311

Hammer TL, Bize P, Saraux C, et al (2022) Repeatability of alert and flight initiation distances in king penguins: effects of colony, approach speed, and weather. Ethology 128:303–16. https://doi.org/10.1111/eth.13264

Hammer TL, Bize P, Gineste B, et al (2023) Disentangling the "many-eyes", "dilution effect", "selfish herd", and "distracted prey" hypotheses in shaping alert and flight initiation distance in a colonial seabird. Behav Process 210:104919. https://doi.org/10.1016/j.beproc.2023.104919

Hess S, Fischer S, Taborsky B (2016) Territorial aggression reduces vigilance but increases aggression towards predators in a cooperatively breeding fish. Anim Behav 113:229–235. https://doi.org/10.1016/j.anbehav.2016.01.008

Hunter S (1991) The impact of avian predator scavengers on king penguin *Aptenodytes patagonicus* chicks at Marion Island. Ibis 133:343–350

Jakobsson S, Brick O, Kullberg C (1995) Escalated fighting behaviour incurs increased predation risk. Anim Behav 49:235–239. https://doi.org/10.1016/0003-3472(95)80172-3

Jouventin P, Dobson FS. Why penguins communicate: the evolution of visual and vocal signals. Academic Press, 2017.

Kleindorfer S, Hoi H, Fessl B (1996) Alarm calls and chick reactions in the moustached warbler, *Acrocephalus melanopogon*. Anim Behav 51:1199–1206

Klvaňová A, Hořáková D, Exnerová A (2011) Nest defence intensity in house sparrows *Passer domesticus* in relation to parental quality and brood value. Acta Ornithol 46:47–54. https://doi.org/10.3161/000164511X589910

Kokko H (1999) Competition for early arrival in migratory birds. J Anim Ecol 68:940–950. https://doi.org/10.1046/j.1365-2656.1999.00343.x

Koskela E, Juutistenaho P, Mappes T, Oksanen TA (2000) Offspring defence in relation to litter size and age: Experiment in the bank vole *Clethrionomys glareolus*. Evol Ecol 14:99–109. https://doi.org/10.1023/A:1011051426666

Lack D (1948) The significance of clutch size. Part III: Some interspecific comparisons. Ibis 90:25–45. https://doi.org/10.1111/j.1474-919X.1948.tb01399.x

Lack D (1947) The Significance of Clutch-size. Part I: Intraspecific Variations. Ibis 89:302–352Le Bohec C, Gauthier-Clerc M, Le Maho Y (2005) The adaptive significance of crèches in the king penguin. Anim Behav 70:527–538. https://doi.org/10.1016/j.anbehav.2004.11.012

Lendvai ÁZ, Chastel O (2008) Experimental mate-removal increases the stress response of female house sparrows: The effects of offspring value? Horm Behav 53:395–401. https://doi.org/10.1016/j.yhbeh.2007.11.011

Lendvai ÁZ, Giraudeau M, Chastel O (2007) Reproduction and modulation of the stress response: An experimental test in the house sparrow. Proc R Soc B Biol Sci 274:391–397. https://doi.org/10.1098/rspb.2006.3735

Lenth R V. (2021) emmeans: Estimated Marginal Means, aka Least-Squares Means. R package version 1.5.5-1. https://CRAN.R-project.org/package=emmeans

Lima SL (2009) Predators and the breeding bird: Behavioral and reproductive flexibility under the risk of predation. Biol Rev 84:485–513. https://doi.org/10.1111/j.1469-185X.2009.00085.x

Lima SL, Dill LM (1990) Behavioral decisions made under the risk of predation: a review and prospectus. Can J Zool 68:619–640. https://doi.org/10.1139/z90-092

Lüdecke D (2024) sjPlot: data visualization for statistics in social sciences. R package version 2.8.16, https://CRAN.R-project.org/package=sjPlot.

Mayer M, Natusch D, Frank S (2019) Water body type and group size affect the flight initiation distance of European waterbirds. PLoS One 14:e0219845. https://doi.org/10.1371/journal.pone.0219845

Mikula P, Díaz M, Albrecht T, et al (2018a) Adjusting risk-taking to the annual cycle of long-distance migratory birds. Sci Rep 8, 13989. https://doi.org/10.1038/s41598-018-32252-1

Mikula P, Díaz M, Møller AP, et al (2018b) Migratory and resident waders differ in risk taking on the wintering grounds. Behav Process 157, 309-314. https://doi.org/10.1016/j.beproc.2018.07.020

Møller AP, Nielsen JT, Garamzegi LZ (2008) Risk taking by singing males. Behav Ecol 19:41–53. https://doi.org/10.1093/beheco/arm098

Montgomerie RD, Weatherhead PJ (1988) Risks and rewards of nest defence by parent birds. Q Rev Biol 63:167–187

Mooring MS, Hart BL (1995) Costs of allogrooming in impala: distraction from vigilance. Anim Behav 49:1414–1416. https://doi.org/10.1006/anbe.1995.0175

Novčić I, Parača V (2022) Seasonal differences in escape behaviour in the urban hooded crow, Corvus cornix. J Vertebr Biol 71:21066. https://doi.org/10.25225/jvb.21066

Olson RS, Hintze A, Dyer FC, et al (2013) Predator confusion is sufficient to evolve swarming behaviour. J R Soc Interface 10:20130305

Olsson O (1996) Seasonal effects of timing and reproduction in the king penguin: a unique breeding cycle. J Avian Biol 27:7–14. https://doi.org/10.2307/3676955

Onnebrink H, Curio E (1991) Brood defense and age of young: a test of the vulnerability hypothesis. Behav Ecol Sociobiol 29:61–68. https://doi.org/10.1007/BF00164296

Pavel V (2006) When do altricial birds reach maximum of their brood defence intensity? J Ethol 24:175–179. https://doi.org/10.1007/s10164-005-0178-y

Petelle MB, McCoy DE, Alejandro V, et al (2013) Development of boldness and docility in yellow-bellied marmots. Anim Behav 86:1147–1154

Petrelli AR, Levenhagen MJ, Wardle R, et al (2017) First to flush: The effects of ambient noise on songbird flight initiation distances and implications for human experiences with nature. Front Ecol Evol 5:67. https://doi.org/10.3389/fevo.2017.00067

Pichegru L, Edwards TB, Dilley BJ, et al (2016) African Penguin tolerance to humans depends on historical exposure at colony level. Bird Conserv Int 26:307–322. https://doi.org/10.1017/S0959270915000313

Piratelli AJ, Favoretto GR, de Almeida Maximiano MF (2015) Factors affecting escape distance in birds. Zoologia 32:438–444. https://doi.org/10.1590/S1984-46702015000600002

Pulliam HR (1973) On the advantages of flocking. J Theor Biol 38:419–422. https://doi.org/10.1016/0022-5193(73)90184-7

Quinn JL, Cresswell W (2006) Testing domains of danger in the selfish herd: sparrowhawks target widely spaced redshanks in flocks. Proc R Soc B Biol Sci 273:2521–2526. https://doi.org/10.1098/rspb.2006.3612

R Development Core Team (2021) A language and environment for statistical computing. R Found Stat Comput

Redmond LJ, Murphy MT, Dolan AC, Sexton K (2009) Parental investment theory and nest defense by eastern kingbirds. Wilson J Ornithol 121:1–11. https://doi.org/10.1676/07-166.1

Roff DA (1992) Evolution of life histories: theory and analysis. Chapman and Hall, New York

Samia DSM, Blumstein DT, Stankowich T, Jr WEC (2016) Fifty years of chasing lizard: new insights advance optimal escape theory. Biol Rev 91:349–366. https://doi.org/10.1111/brv.12173

Sandercock BK (1994) The effect of manipulated brood size on parental defence in a precocial bird, the Willow Ptarmigan. J Avian Biol 25:281–286. https://doi.org/10.2307/3677275

Samia DS, Blumstein DT, Díaz M, et al (2017) Rural-urban differences in escape behavior of European birds across a latitudinal gradient. Front Ecol Evol 5:66.

Schielzeth H, Dingemanse NJ, Nakagawa S, et al (2020) Robustness of linear mixed-effects models to violations of distributional assumptions. Methods Ecol Evol 11:1141–1152. https://doi.org/10.1111/2041-210X.13434

Schmid B, Tam-Dafond L, Jenni-Eiermann S, et al (2013) Modulation of the adrenocortical response to acute stress with respect to brood value, reproductive success and survival in the Eurasian hoopoe. Oecologia 173:33–44. https://doi.org/10.1007/s00442-013-2598-7

Schull Q, Dobson FS, Stier A, et al (2016) Beak color dynamically signals changes in fasting status and parasite loads in king penguins. Behav Ecol 27: 1684-1693.

Schull Q, Robin J-P, Dobson FS, et al (2018) Experimental stress during molt suggests the evolution of condition-dependent and condition-independent ornaments in the king penguin. Ecol Evol 8: 1084-1095.

Smith-Castro JR, Rodewald AD (2010) Behavioral responses of nesting birds to human disturbance along recreational trails. J F Ornithol 81:130–138. https://doi.org/10.1111/j.1557-9263.2010.00270.x

Stankowich T, Blumstein DT (2005) Fear in animals: a meta-analysis and review of risk assessment. Proc R Soc B Biol Sci 272:2627–2634. https://doi.org/10.1098/rspb.2005.3251

Stearns SC (1992) The evolution of life histories. Oxford University Press, Oxford

Stonehouse B (1960) The king penguin *Aptenodytes patagonica* of South Georgia. Falkland Islands Dependencies Survey. Scientific Reports No. 23.

Stier A, Viblanc VA, Massemin-Challet S, Handrich Y, Zahn S, Rojas ER, Saraux C, Le Vaillant M, Prud'Homme O, Grosbellet E, Robin JP (2014) Starting with a handicap: phenotypic differences between early-and late-born king penguin chicks and their survival correlates. Funct Ecol 28:601-11. https://doi.org/10.1111/1365-2435.12204

Strnadová I, Němec M, Strnad M, et al (2018) The nest defence by the red-backed shrike (*Lanius collurio*) - support for the vulnerability hypothesis. J Avian Biol 49:jav-01726. https://doi.org/10.1111/jav.01726

Svagelj WS, Magdalena Trivellini M, Quintana F (2012) Parental investment theory and nest defence by imperial shags: effects of offspring number, offspring age, laying date and parent sex. Ethology 118:251–259. https://doi.org/10.1111/j.1439-0310.2011.02003.x

Thünken T, Meuthen D, Bakker TCM, Kullmann H (2010) Parental investment in relation to offspring quality in the biparental cichlid fish Pelvicachromis taeniatus. Anim Behav 80:69–74. https://doi.org/10.1016/j.anbehav.2010.04.001

Tjur T (2009) Coefficients of determination in logistic regression models - A new proposal: The coefficient of discrimination. Am Stat 63: 366-372. https://doi.org/10.1198/tast.2009.08210

Trivers RL (1972) Parental investment and sexual selection. In: Campbell B (ed) Sexual selection and the descent of man 1871-1971. pp 136–179

Tryjanowski P, Goławski A (2004) Sex differences in nest defence by the red-backed shrike *Lanius collurio*: Effects of offspring age, brood size, and stage of breeding season. J Ethol 22:13–16. https://doi.org/10.1007/s10164-003-0096-9

Tseng SP, Lin YY, Yang YC, et al (2017) Injury feigning in the Savanna nightjar: a test of the vulnerability and brood value hypotheses. J Ornithol 158:507–516. https://doi.org/10.1007/s10336-016-1400-0

Van Heezik YM, Seddon PJ, Cooper J, Plös AL (1994) Interrelationships between breeding frequency, timing and outcome in King Penguins *Aptenodytes patagonicus*: are King Penguins biennial breeders? Ibis 136:279–284. https://doi.org/10.1111/j.1474-919X.1994.tb01096.x

Ventura SPR, Galdino CAB, Peixoto PEC (2021) Fatal attraction: territorial males of a neotropical lizard increase predation risk when females are sexually receptive. Behav Ecol Sociobiol 75:170. https://doi.org/10.1007/s00265-021-03112-2

Viblanc VA, Dobson FS, Stier A, et al. (2016) Mutually honest? Physiological 'qualities' signalled by colour ornaments in monomorphic king penguins. Biol J Linn Soc 118: 200-214.

Viblanc VA, Gineste B, Robin J, Groscolas R (2016) Breeding status affects the hormonal and metabolic response to acute stress in a long-lived seabird, the king penguin. Gen Comp Endocrinol 236:139–145. https://doi.org/10.1016/j.ygcen.2016.07.021

Viblanc VA, Smith AD, Gineste B, et al (2015) Modulation of heart rate response to acute stressors throughout the breeding season in the king penguin *Aptenodytes patagonicus*. J Exp Biol 218:1686–1692. https://doi.org/10.1242/jeb.112003

Viera VM, Viblanc VA, Filippi-Codaccioni O, et al (2011) Active territory defence at a low energy cost in a colonial seabird. Anim Behav 82:69–76. https://doi.org/10.1016/j.anbehav.2011.04.001

Villanueva C, Walker BG, Bertellotti M (2014) Seasonal variation in the physiological and behavioral responses to tourist visitation in Magellanic penguins. J Wildl Manage 78:1466–1476. https://doi.org/10.1002/jwmg.791

Weatherhead PJ (1979) Do savannah sparrows commit the concorde fallacy? Behav Ecol Sociobiol 5:373–381. https://doi.org/10.1007/BF00292525

Weimerskirch H, Stahl JC, Jouventin P (1992) The breeding biology and population dynamics of King Penguins *Aptenodytes patagonica* on the Crozet Islands. Ibis 134:107–117. https://doi.org/10.1111/j.1474-919X.1992.tb08387.x

Williams CT, Lane JE, Humphries MM, et al (2014) Reproductive phenology of a food-hoarding mast-seed consumer: resource- and density-dependent benefits of early breeding in red squirrels. Oecologia 174:777–788. https://doi.org/10.1007/s00442-013-2826-1

Williams GC (1966) Natural selection, the costs of reproduction, and a refinement of Lack's principle. Am Nat 100:687–690

Wilson RP, Culik B, Danfeld R, Adelung D (1991) People in Antarctica - how much do Adélie Penguins *Pygoscelis adeliae* care? Polar Biol 11:363–370. https://doi.org/10.1007/BF00239688

Ydenberg RC, Dill LM (1986) The economics of fleeing from predators. Adv Study Behav 16:229–249. https://doi.org/10.1016/S0065-3454(08)60192-8

Yee J, Lee J, Desowitz A, Blumstein DT (2013) The costs of Conspecifics: are social distractions or environmental distractions more salient? Ethology 119:480–488. https://doi.org/10.1111/eth.12085

## FIGURE CAPTIONS

**Fig 1. Effect of life history stage on Alert Distance (AD) in king penguins.** (Left panel) Marginal means along with their 95% CI are presented in black, raw data is overlaid in red. Groups not sharing similar superscripts (bottom) are significantly different for P<0.05 (Tukey HSD). (Right panel) Standardized linear mixed model estimates (z-scores) and 95% confidence for life

history stage (non-breeding vs. courting vs. settled pairs vs. incubating vs. brooding (TDC or TIC) birds) and controlling variables on AD (see text). Molting birds were taken as a reference category. Negative effects are indicated in red, positive effects in blue.

**Fig 2. Effect of life history stage on the probability to flee in king penguins.** (Left panel) Marginal means from a binomial GLM along with their 95% CI are presented in black, raw data is overlaid in red. Groups not sharing similar superscripts are significantly different for P<0.05 (Tukey HSD). (Right panel) Odds ratios and 95% confidence for life history stage (brooding TDC vs. brooding TIC) and controlling variables on the probability to flee (see text). Incubating birds were taken as a reference category. Negative effects are indicated in red, positive effects in blue.

**Fig 3. Effect of life history stage on Flight Initiation Distance (FID, log-transformed) in king penguins.** (Left panel) Marginal means along with their 95% CI are presented in black, raw data is overlaid in red. Groups not sharing similar superscripts (bottom) are significantly different for P<0.05 (Tukey HSD). (Right panel) Standardized linear mixed model estimates (z-scores) and 95% confidence for life history stage (non-breeding vs. courting vs. settled pairs vs. incubating vs. brooding (TDC or TIC) birds) and controlling variables on FID (see text). Molting birds were taken as a reference category. Negative effects are indicated in red, positive effects in blue.

**Fig 4. Effect of life history stage on Flight Distance (FD, log-transformed) in king penguins.** (Left panel) Marginal means along with their 95% CI are presented in black, raw data is overlaid in red. Groups not sharing similar superscripts (bottom) are significantly different for P<0.05

(Tukey HSD). (Right panel) Standardized linear mixed model estimates (z-scores) and 95% confidence for life history stage (non-breeding vs. courting vs. settled pairs vs. incubating vs. brooding (TDC or TIC) birds) and controlling variables on FD (see text). Molting birds were taken as a reference category. Negative effects are indicated in red, positive effects in blue.

**Fig 5. Interactive effects of weighted aggression towards the experimenter and breeding stage on bird's fleeing probability in king penguins.** All birds sampled were either incubators or brooders, groups for which at least 1 aggression directed towards the experimenter was recorded. The predicted values and 95%CI are shown (GLMM, binomial error distribution) controlling for life history stage (incubator vs. brooder of young vs. brooder of older chicks). The raw binomial data for a given bird are either to flee (1) or not (0), however some jitter along the x and y axes has been artificially added to the plot in order to better visualize individual data points.

**FIGURES**

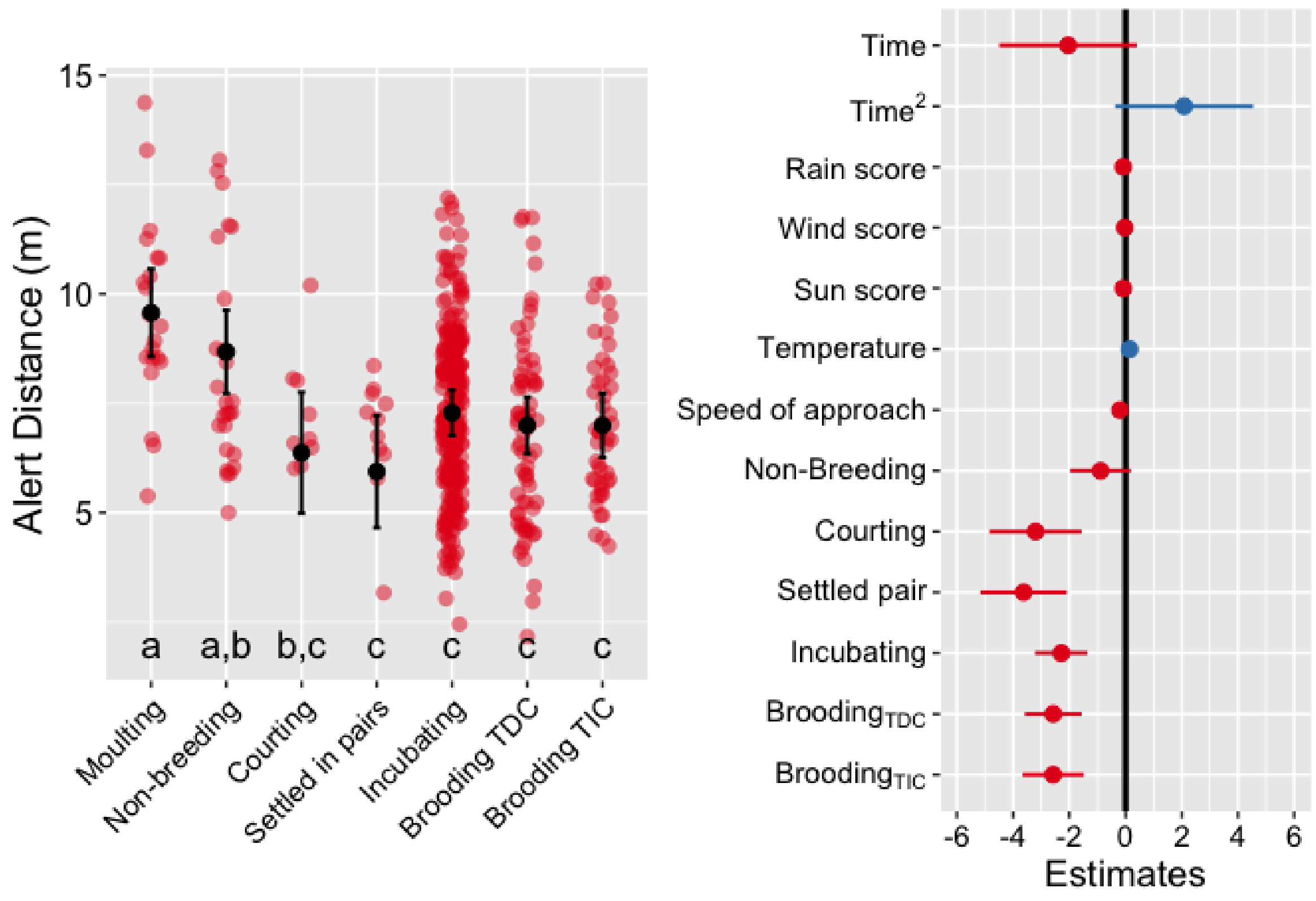

**Fig 1.**

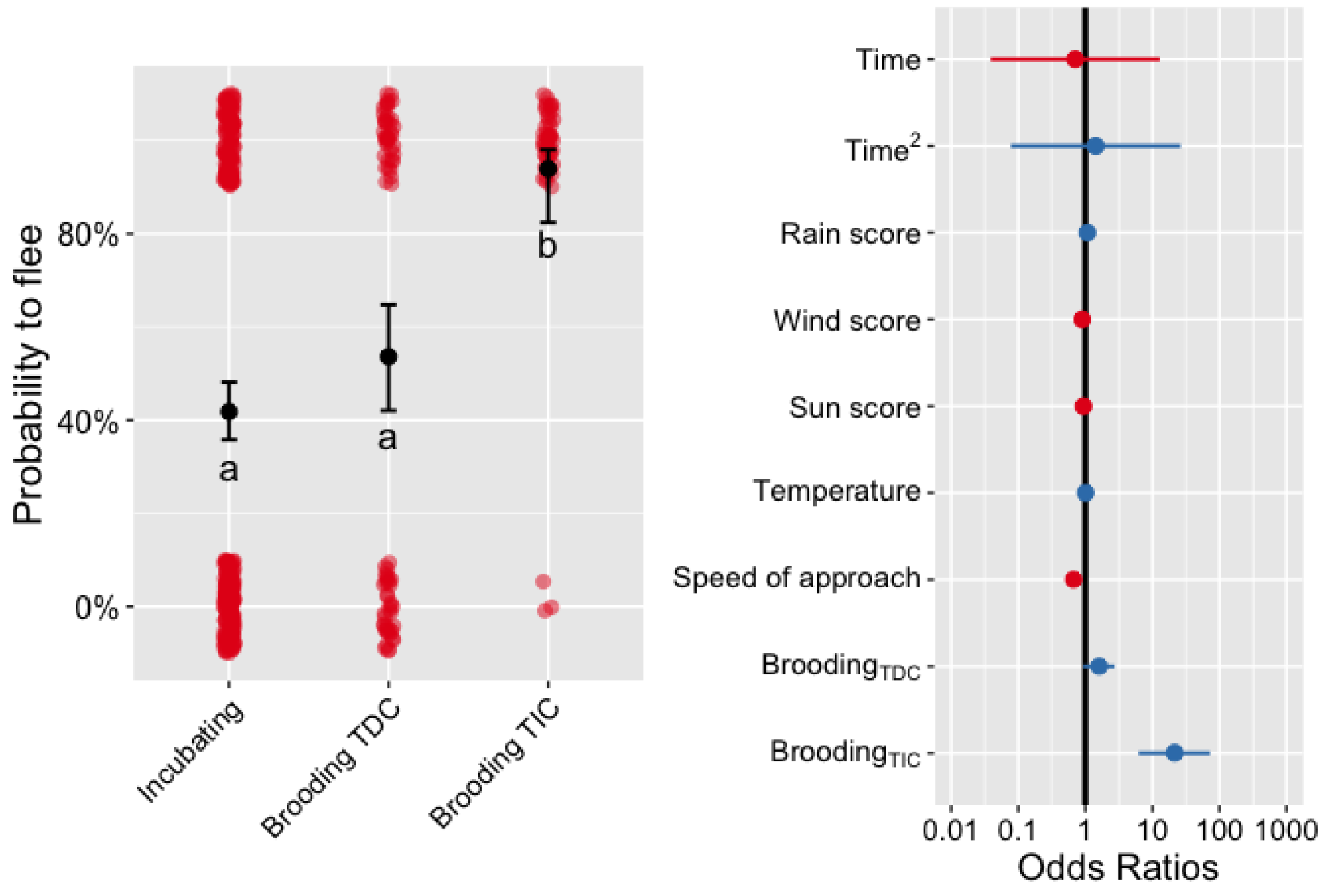

**Fig 2.**

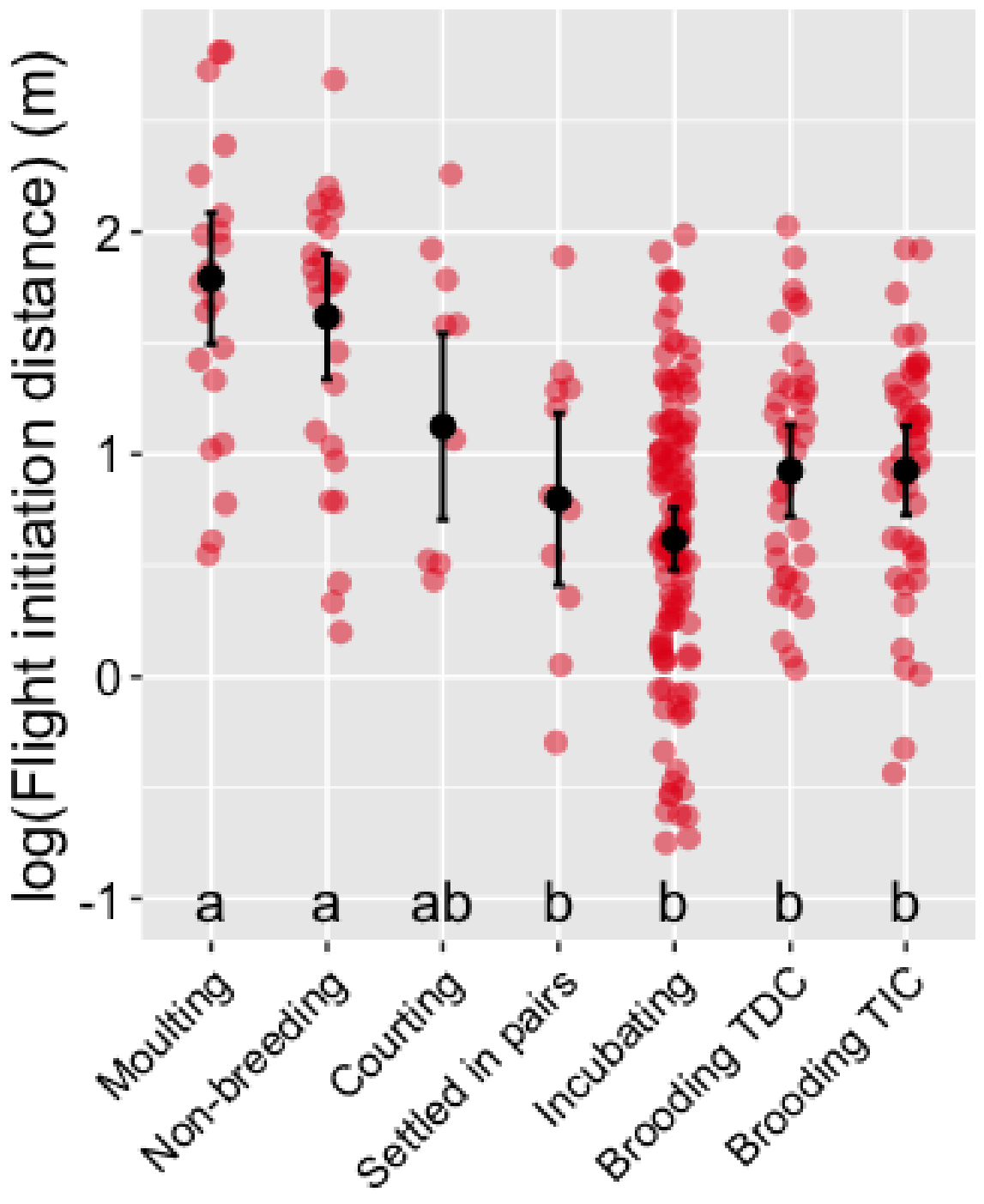

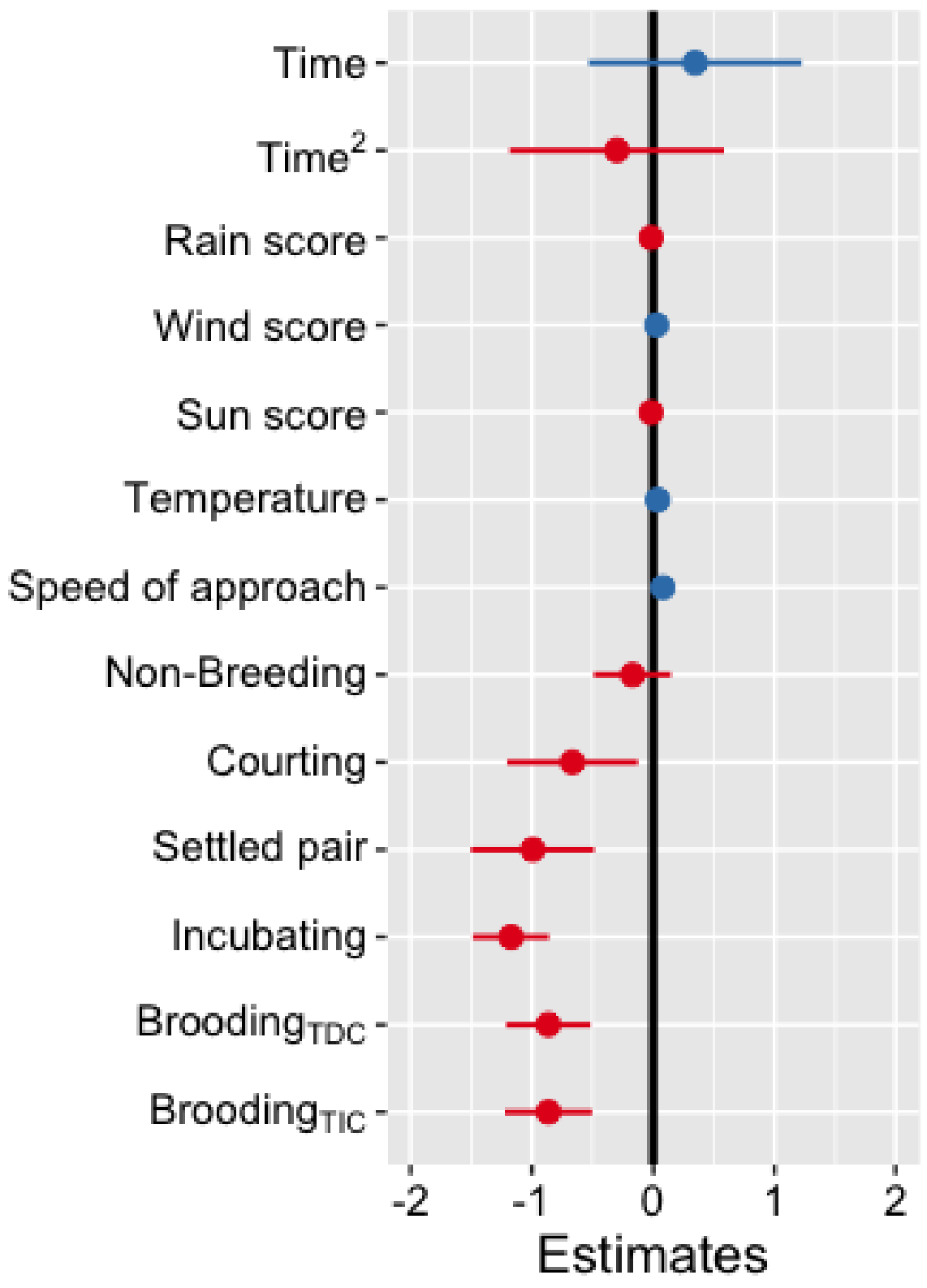

**Fig 3.**

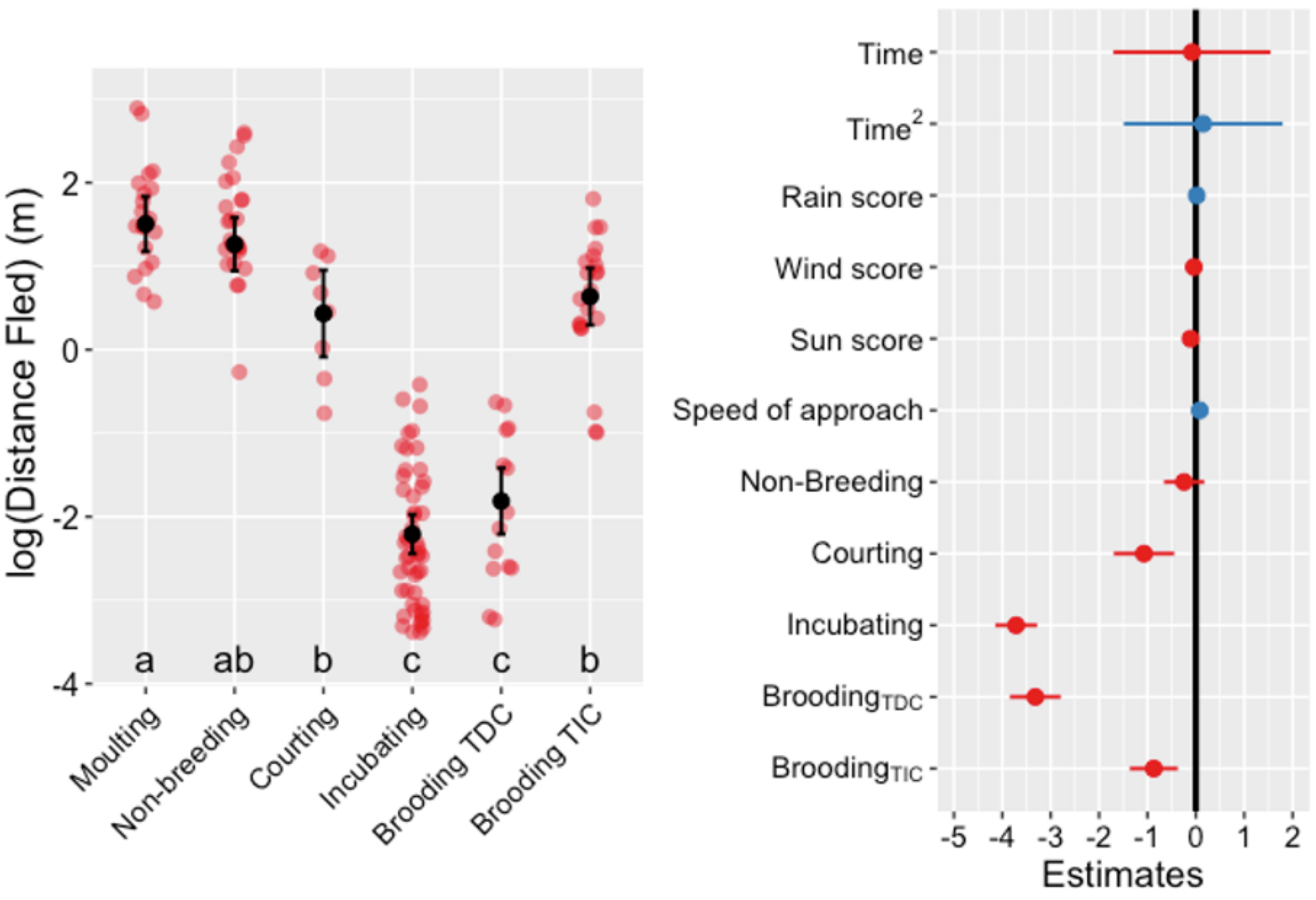

**Fig 4.**

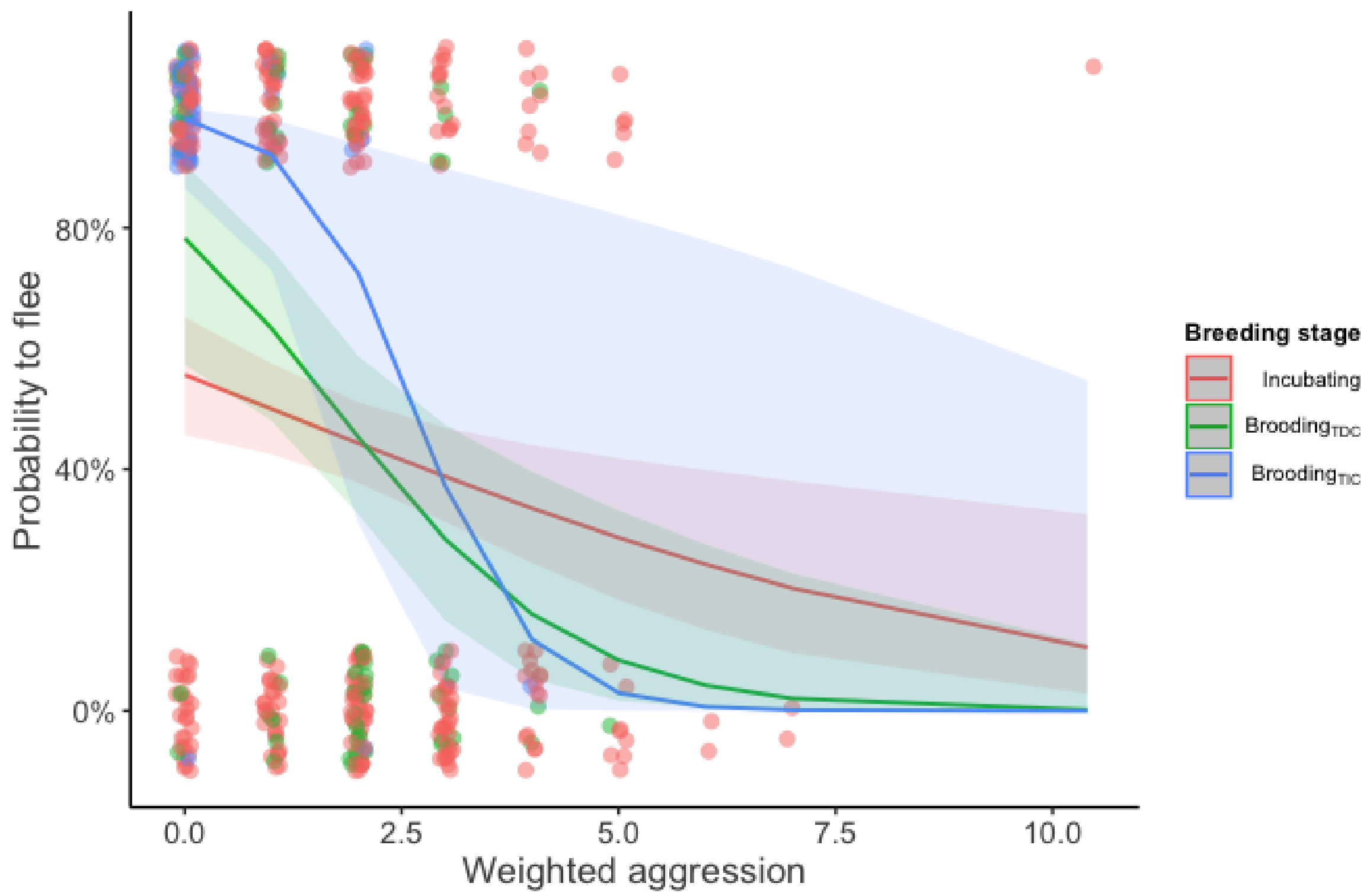

**Fig 5.**